\documentclass[sigconf,nonacm]{acmart}

\usepackage{listings}
\usepackage{graphicx}
\usepackage{enumitem}

\setlist[itemize]{leftmargin=1em, topsep=4pt}

\title{Translating PL/I Macro Procedures into Java Using Automatic Templatization and Large Language Models}

\author{Takaaki Tateishi}
\affiliation{
  \institution{IBM Research}
  \city{Tokyo}
  \country{Japan}   
}
\email{tate@jp.ibm.com}
\author{Yasuharu Katsuno}
\affiliation{
  \institution{IBM Research}
  \city{Tokyo}
  \country{Japan}   
}
\email{katsuno@jp.ibm.com}

\begin{document}


\begin{abstract}
Modernizing legacy enterprise systems often involves translating PL/I programs into modern languages
such as Java.
This task becomes significantly more complex when PL/I macro procedures are involved.
The PL/I macro procedures are considered string-manipulating programs that generate PL/I code, and they make automated translation more complex.
Recently, large language models (LLMs) have been explored for automated code translation.
However, LLM-based code translation struggles to translate the PL/I macro procedures
to Java programs that reproduce the behavior of the plain PL/I code generated
by the original PL/I macro procedures.

This paper proposes a novel method called templatization, which uses symbolic execution
to generate code templates (code with named placeholders) as an intermediate representation.
In this approach, symbolic values are treated as parts of macro-generated code.
By symbolically executing macro procedures and generating code templates,
our approach facilitates LLMs to generate readable and maintainable Java code.
Our preliminary experiment on ten PL/I macro procedures shows that the LLM-based translation
through templatization successfully generates Java programs
that reproduce the behavior of the macro-generated PL/I programs.
\end{abstract}

\maketitle

\section*{Introduction}
\label{sec:intro}
\setcounter{section}{1}

Legacy programs written using PL/I are still widely used in enterprise systems.
However, modernizing these systems often involves translating PL/I programs into a modern language like Java.
Recently, approaches using machine learning and pretrained large language models (LLMs) have been actively explored for code translation tasks~\cite{CodeXGLUE,CodeTransOcean}.
The syntax of PL/I closely resembles that of modern procedural languages.
Our preliminary investigation confirms that many recent LLMs are capable of translating plain PL/I programs,
which are composed solely of basic control statements (e.g., conditionals and procedure calls)
and arithmetic expressions, into equivalent Java programs.

However, a significant challenge arises when PL/I programs include macro procedures.
(1) PL/I macro procedures can be regarded as string-manipulating programs that generate PL/I programs.
Unlike PL/I, Java lacks a code-generation mechanism similar to PL/I macro procedures.
(2) These macro procedures are written using PL/I, and the strings manipulated
within them correspond to the PL/I programs to be generated.
To the best of our knowledge, no existing papers have addressed such programs.
(3) If the concrete string values passed to macro procedures are known, macro expansion
can be performed by interpreting the macro instructions, resulting in plain PL/I code.
However, macro procedures are typically reused across multiple PL/I programs,
and we obtain multiple macro-expanded PL/I programs from a single macro procedure.
When these are translated into Java, they result in multiple Java programs,
which compromises the original modularity.

\begin{figure*}[t]
\includegraphics[width=440pt]{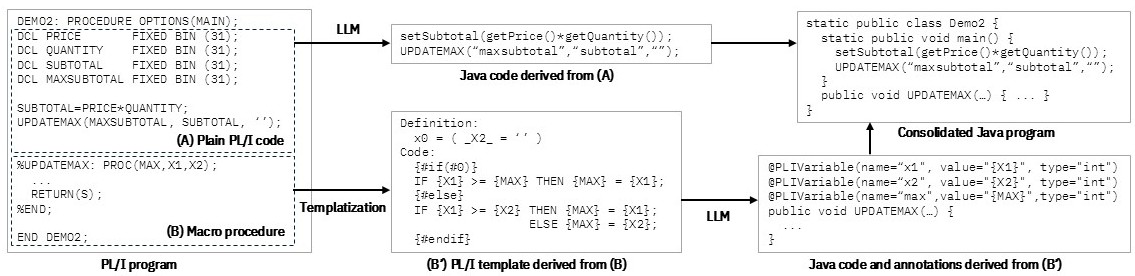}
\caption{Overview of Translation from PL/I to Java}
\label{fig:overview}
\end{figure*}

\noindent
\textbf{Our Approach: }
Given the above challenges, we employ the translation process illustrated in Figure \ref{fig:overview}.
This pipeline translates PL/I to Java through two separate paths:
one for plain PL/I code and another for PL/I macro procedures.
In this paper, we focus on the path for PL/I macro procedures, as plain PL/I code can be directly
translated into Java by an LLM.
For PL/I macro procedures, we first translate them to PL/I templates that represent the
code generated by the macro procedures, which we refer to as \emph{templatization}.
This templatization process is performed through symbolic execution tailored for handling
code-generating code.
Each resulting template consists of two components: a definition part and a code part.
The definition parts of the templates are translated into Java using straightforward
rules that make use of the execution traces obtained through symbolic execution,
the details of which are omitted in this paper.
The code parts of the templates are translated into Java code with annotations
using the same LLM and a prompt extended to handle placeholders in the PL/I templates.
These annotations simulate the behavior of PL/I macro procedures.
This approach preserves the modularity and maintainability of the original code,
enabling the LLM to generate Java programs that reproduce the behavior of the macro-generated PL/I programs.

\noindent
\textbf{Contributions:} 
The contributions of this paper are as follows:
(1) PL/I template generation through symbolic execution tailored for PL/I macro processing.
To the best of our knowledge, ours is the first application of symbolic execution used in this manner.
(2) Java annotations that implement the parameterized variable names to accommodate variable names
generated by the PL/I macro procedures.

\noindent
\textbf{Organization: }
This paper is organized as follows.
Section \ref{sec:related} introduces related work.
Section \ref{sec:method} describes our proposed method.
Section \ref{sec:experiment} demonstrates the effectiveness of templatization through experiments.
Finally, Section \ref{sec:plans} outlines our future plans, and Section \ref{sec:conclusion} concludes this paper.

\section{Related Work}
\label{sec:related}

Recent advancements in LLMs has resulted in growing interest in automated code translation,
especially for legacy programming languages such as COBOL~\cite{CodeTransOcean}
and FORTRAN~\cite{FortranToCPP}.
Several recent approaches enhance LLM-based code translation by integrating
external sources of semantic or structural information as we leverage symbolic execution and code template.
CoTrans~\cite{CoTran} incorporates compiler feedback and symbolic execution during fine-tuning,
allowing the model to learn from program behavior and improve accuracy.
INTERTRANS~\cite{INTERTRANS} uses intermediate programming languages as semantic bridges to harness the multilingual capabilities of code LLMs.
LLML{\scriptsize IFT}~\cite{llmlift} combines LLMs with formal verification, which uses a symbolic solver,
to ensure that translated code preserves the original program's semantics and correctness.

Collectively, these studies demonstrate the growing capabilities of LLMs in code translation.
Still, there are challenges in handling code that generates other code, such as macro procedures,
which require specialized techniques like our templatization.

\section{LLM-based Translation Through Templatization}
\label{sec:method}

\begin{figure}
\begin{lstlisting}
%UPDATEMAX: PROC(PREFIX,MAX,X1,X2) RETURNS(CHAR);
 DCL PREFIX CHAR;
 DCL MAX CHAR;
 DCL X1 CHAR;
 DCL X2 CHAR;
 DCL S CHAR;
 S = '';
 IF X2 = '' THEN
   DO;
   S = S||'IF '||PREFIX||X1||' >= '||PREFIX||MAX||' THEN ';
   S = S||'   '||PREFIX||MAX||' = '||PREFIX||X1||'; ';
   END;
 ELSE
   DO;
   S = S||'IF '||PREFIX||X1||' >= '||PREFIX||X2||' THEN ';
   S = S||'   '||PREFIX||MAX||' = '||PREFIX||X1||'; ';
   S = S||'ELSE ';
   S = S||'   '||PREFIX||MAX||' = '||PREFIX||X2||'; ';
   END;
 RETURN(S);
%END;
\end{lstlisting}
\vspace{-4pt}
\caption{PL/I macro procedure}
\label{fig:macroproc}
\vspace{-8pt}
\end{figure}

\begin{figure}[t]
\begin{lstlisting}
@PLIVariable(name="prefixX1", value="{_PREFIX_}{_X1_}", type="int")
@PLIVariable(name="prefixX2", value="{_PREFIX_}{_X2_}", type="int")
@PLIVariable(name="prefixMax", value="{_PREFIX_}{_MAX_}", type="int")
public void targetMethod(String _PREFIX_, String _MAX_,
                         String _X1_, String _X2_) {
  boolean x0 = (_X2_ = '');
  if(x0) {
    if (getPrefixX1() >= getPrefixMax()) {
      setPrefixMax(getPrefixX1());
    }
  } else {
    if (getPrefixX1() >= getPrefixX2()) {
      setPrefixMax(getPrefixX1());
    } else {
      setPrefixMax(getPrefixX2());
    }
  }
}
\end{lstlisting}
\vspace{-4pt}
\caption{Generated Java code}
\label{fig:generated}
\vspace{-8pt}
\end{figure}

In this section, we first describe the syntax of the PL/I macro procedures targeted for our approach
and the structure of the expected Java code generated by the LLM-based translation
including our custom Java annotations.
We then explain the details of the templatization process.
Finally, we present the prompt design used to guide the LLMs in generating such annotated Java code.

\subsection{PL/I Macro Procedures and Corresponding Java Methods}

PL/I macro language has a syntax similar to that of the PL/I language,
but the values are limited to strings and numbers.
Figure~\ref{fig:macroproc} shows an example of a PL/I macro procedure that generates PL/I code
to update a maximum value.
Note that this macro procedure takes 4 parameters: \texttt{PREFIX}, \texttt{MAX}, \texttt{X1}, and \texttt{X2}.
Each parameter represents a string used to construct variable names.

A simplified subset of the syntax of PL/I macro procedures~\cite{ibm2022pli} is defined as follows:
\\
$
\begin{array}{ll}
\text{expression}& e ::= x \ | \ s \ | \ n \\
                 & \hspace{4ex} |\ e_1 \texttt{||} e_2 \ |\ e_1 \texttt{=} e_2
                   \ |\ \texttt{F}(e_1,\cdots) \ |\ \cdots \\
\text{statement} & \textit{stmt} ::= x \texttt{=} e \texttt{;} \\
                 & \hspace{4ex}    |\ \texttt{IF } e \texttt{ THEN } \textit{stmts} \texttt{ ELSE } \textit{stmts} \texttt{ END} \\
                 & \hspace{4ex}    |\ \texttt{DO WHILE } e \texttt{;} \textit{stmts} \texttt{ END }\\
                 & \hspace{4ex}    |\ \texttt{DO } \textit{stmts} \texttt{ END } \\
\text{statements}& \textit{stmts} ::= \textit{stmt} \texttt{;} \cdots \texttt{;} \textit{stmt} \texttt{;} \\
\text{procedure} & \textit{mproc} ::= \\
                 & \hspace{2ex} \texttt{\%} \textit{name} \texttt{: PROC} \ (x_1,x_2,\cdots) \texttt{ RETURNS CHAR; } \\
                 & \hspace{4ex}             \textit{stmts} \ \texttt{RETURN}(e) \texttt{;} \texttt{ END }\\
\end{array}
$
\\
Expressions $e$ consist of variables $x$, strings $s$, and numbers $n$.
They may also include concatenation $e_1 \,\texttt{||}\, e_2$ and
other string operations $\texttt{F}(e_1,\cdots)$
such as substring extraction, where the subscripted $e$ also represents an expression.
Statements $\textit{stmt}$ include assignments, conditional branches, while-loops, and statement blocks,
where a sequence of statements $\textit{stmts}$ consists of multiple statements separated by semicolons.
A macro procedure named \textit{name} is defined using \texttt{\%} and \texttt{PROC}.

We expect that the Java methods generated from the PL/I macro procedures by an LLM reproduce the behavior
of the original macro-generated PL/I programs.
Therefore, the generated Java methods are designed to take parameters of type \texttt{String},
which serve as meta-variables similar to those in the original macro procedures,
as shown in Figure \ref{fig:generated}.
These meta-variables are used to determine the actual variable names in the generated code,
and the resolved names are then used to obtain the corresponding values.
To simplify the retrieval of the values through these meta-variables,
we introduce the annotation \texttt{@PLIVariable(name,value,type)}, which automatically defines
corresponding setter and getter methods.

An annotation processor interprets such annotations and generate setter and getter methods
that maintain the relationship between variable names, values, and types using a key-value store.
For example, the following is the implementation of the getter method \texttt{getPrefixX1()},
which corresponds to the annotation
\\
\texttt{@PLIVariable(name="prefixX1", value="\{\_PREFIX\_\}\{\_X1\_\}", type="int")} shown in Figure~\ref{fig:generated}:
\begin{lstlisting}
public int getPrefixX1() {
  String varName = this.metavar.get("PREFIX") + this.metavar.get("X1");
  Object obj = this.kv.get(varName);
  return (Integer)obj;
}
\end{lstlisting},
where the variable \texttt{varName} is assigned the concatenated string that is represented by
\texttt{"\{\_PREFIX\_\}\{\_X1\_\}"} and consists of the values of \texttt{PREFIX} and \texttt{X1}.
Using this computed name, the method retrieves the corresponding value from the key-value store
\texttt{this.kv}, performs the appropriate type conversion, and returns the result.

\subsection{Templatization Based on Symbolic Execution}

\begin{figure}[t]
\begin{lstlisting}
Symbolic values:
 x0 = (X2 = '')

Return values:
 {#if(#0)}
 IF {_PREFIX_}{_X1_} >= {_PREFIX_}{_MAX_}
 THEN    {_PREFIX_}{_MAX_} = {_PREFIX_}{_X1_};
 {#else}
 IF {_PREFIX_}{_X1_} >= {_PREFIX_}{_X2_}
 THEN    {_PREFIX_}{_MAX_} = {_PREFIX_}{_X1_};
 ELSE    {_PREFIX_}{_MAX_} = {_PREFIX_}{_X2_};
 {#endif}
\end{lstlisting}
\caption{PL/I template}
\label{fig:template}
\end{figure}

Figure~\ref{fig:template} shows the template generated from the macro procedure in Figure~\ref{fig:macroproc} through symbolic execution, which is referred to as templatization.
In our symbolic execution, a symbolic value $v$ is defined as follows:
\\
$
\begin{array}{rlr}
v ::= & [w_1, \ldots, w_n] & \text{(symbolic value)} \\
    | & (\texttt{"call"}, F, v_1, v_2, \ldots ) & \text{(function call)} \\
w ::= &  s                 & \text{(string literal)} \\
     | & x                 & \text{(variable)} \\
     | & (\texttt{"if"}, x ) \ |\ (\texttt{"else"}) \ |\ (\texttt{"endif"})
                           & \text{(conditional constructs)} \\
\end{array}
$
\\
Intuitively, a symbolic value abstracts macro-generated code, which consists of code fragments represented as text.
Conditional constructs allow two distinct code fragments, denoted by symbolic values $v_1$ and $v_2$, to be encoded as a single symbolic value:
$[(\texttt{"if"},x_3)] + v_1 + [(\texttt{"else"})] + v_2 + [(\texttt{"endif"})]$,
where $x_3$ is a conditional variable, and $+$ represents list concatenation.

Given a sequence of statements $\textit{stmt}_1; \cdots; \textit{stmt}_n$
and variables \( x_1, \cdots, x_m \) as parameters of a macro procedure,
we derive a template from the symbolic state \( \sigma' \), which is obtained
through symbolic execution
$\sigma' = \theta(\textit{stmt}_1; \cdots; \textit{stmt}_n, \sigma)$,
where the initial state is defined as
$\sigma = \{x_1 \mapsto x_1, \cdots, x_m \mapsto x_m\}$.

The resulting symbolic state $\sigma'$ constitutes the definition part of the template. The code part of the template is constructed by extracting the symbolic value associated with the return variable.
For example, the symbolic value
$[(\texttt{"if"}, x_3), x_1, (\texttt{"else"}), x_2, (\texttt{"endif"})]$
is encoded in the template as
\verb|{#if(#3)}{#1}{#else}{#2}{#endif}|,
where the symbolic values are enclosed in braces and prefixed
with \texttt{\#} to indicate placeholders,
except for the symbolic values associated with the parameters of the macro procedures,
which retain their original names (e.g., \verb|{_MAX_}|).

To define symbolic execution $\theta$, 
we further simplify the PL/I macro language to the following types of statements:
\\
\begin{tabular}[b]{l}
\texttt{$x$ = $s$} \hfill (string substitution) \\
\texttt{$x$ = $x_1$ || $x_2$} \hfill (concatenation) \\
\texttt{$x$ = $F$($x_1$, $\cdots$)} \hfill (string operation) \\
\texttt{IF $x$ THEN \textit{stmts} ELSE \textit{stmts} END} \hfill (conditional branch) \\
\end{tabular}\\
, where $x$, $s$, $F$ and \textit{stmts} represent variables, string values, function symbols, and lists of statements, respectively.
Note that loops are assumed to be unrolled when the number of iterations is fixed.
If the loop condition depends on symbolic values, we unroll the loop for a fixed number of iterations by replacing the loop condition with conditional statements.
We then rely on the downstream LLM-based translation to merge and simplify the resulting code.

The state update rules for $\sigma' = \theta(\textit{stmt}, \sigma)$ are defined as follows.
\begin{itemize}
  \item \textbf{Assignment:}  
  If $\mathit{stmt} = x \texttt{=} s$,
  then $\sigma' = \sigma[x \mapsto s]$.

  \item \textbf{Concatenation:}  
  If $\mathit{stmt} = x \texttt{=} x_1 \texttt{||} x_2$,
  then $\sigma' = \sigma[x \mapsto \sigma(x_1) + \sigma(x_2)]$.
    
  \item \textbf{Function call} (e.g., substring):  
  If $\mathit{stmt} = x \texttt{=} F(x_1, \cdots)$, then\\
  $
    \sigma' =
    \begin{cases}
    \sigma[x \mapsto [F(\sigma(x_1), \cdots)]] & \text{if all } \sigma(x_i) \text{ are concrete} \\
    \sigma[x \mapsto [x'],                     & \text{otherwise} \\
    \quad  x' \mapsto (\texttt{"call"}, F, \sigma(x_1), \cdots)] & \text{$x'$ is a fresh variable} \\
    \end{cases}
  $
  
  \item \textbf{Conditional statement:}  
  If \textit{stmt} $=$ \texttt{IF} $x$ \texttt{THEN} \textit{then-stmts} \texttt{ELSE} \textit{else-stmts} \texttt{END},
  then symbolic execution proceeds as follows:
  \begin{itemize}[topsep=0pt]
    \item we symbolically execute \textit{then-stmts} and \textit{else-stmts} and obtain 
          $\sigma' = \theta(\textit{then-stmts}, \sigma)$,
          and $\sigma' = \theta(\textit{else-stmts}, \sigma)$.
    \item For every variable $x'$ in $\sigma'$, $x'$ is updated only in \textit{then-stmts}, \\
    then $\sigma' := \sigma[x' \mapsto v']$,\\
    where $v' = [\texttt{("if", $x$)}] + \sigma'(x) + [\texttt{("endif")}]$

    \item If $x'$ is updated only in \textit{else-stmts}, \\
    then $\sigma' := \sigma[x' \mapsto v']$,\\
    where $v' = [\texttt{("if", $x$)}, \texttt{("else")}] + \sigma''(x) + [\texttt{("endif")}]$

    \item If $x'$ is updated in both branches,\\
    then $\sigma' := \sigma[x' \mapsto x']$, where $v' = [\texttt{("if", $x$)}] + \sigma'(x) + [\texttt{("else")}] + \sigma''(x) + [\texttt{("endif")}]$
  \end{itemize}
\end{itemize}
Here, $v_1 + v_2$ represents the list concatenation of two symbolic values $v_1$ and $v_2$.
A fresh variable $x'$ is introduced to represent the result of a function call whose arguments are not all concrete, and to prevent the function call from appearing in a template.
We naturally extend $\theta$ to sequence of statements as
$\theta(\textit{stmt}_1;\cdots;\textit{stmt}_n,\sigma)
= \theta(\textit{stmt}_{n}, ...\theta(\textit{stmt}_1,\sigma))$.

\subsection{LLM Translation}

\begin{figure}
\begin{lstlisting}
# Example of Translation From PL/I to Java
## PL/I:
/* MP: BEGIN TEAMPLTE */
/* comment */
A{_BAR_}B = X{_FOO_}Y + 1;
/* MP: END TEMPLATE */
## Java and Annoations:
@PLIVariable(name="aBarB", value="A{_BAR_}B", type="int")
@PLIVariable(name="xFooY", value="X{_FOO_}Y", type="int")
public void targetMethod() {
  this.log("/* comment */");
  setABarB(getXFooY() + 1);
}
... insert as many examples as necessary here ...

# Question:
Translate the following PL/I code into Java according to the above examples.
Each consecutive placeholder (e.g., {_X_}{_Y_}) represents a single variable.
## Answer Format:
* The resulting Java code should be enclosed by "```java" and "```end".
# Answer:
\end{lstlisting}
\vspace{-5pt}
\caption{The prompt for translating PL/I template into Java}
\label{fig:prompt}
\vspace{-10pt}
\end{figure}

Figure~\ref{fig:prompt} illustrates the prompt design used for translating PL/I templates
into Java code using an LLM.
Lines 1--13 present an in-context example that demonstrates the correspondence between
a PL/I template and its translated Java code.
This example considers the following patterns:
(1) Construction patterns of PL/I variable names using meta-variables,
(2) Usage patterns of variables, and
(3) Comment patterns.
Additional in-context examples are inserted to cover other patterns in a similar manner.
Lines 17--21 provide explicit instructions to the LLM.
To programmatically extract the generated Java code from the LLM response,
the prompt instructs the LLM to enclose the resulting Java code between the markers
``\verb|```java|'' and ``\verb|```end|''.

\section{Experiment}
\label{sec:experiment}

To evaluate the effectiveness of the method, we created 10 realistic PL/I macro procedures,
comprising a total of 146 lines of code with 15 conditional branches and 5 built-in functions such as \texttt{SUBSTR}, including the one shown in Figure~\ref{fig:macroproc}.
Our implementation of symbolic execution for PL/I macro procedures successfully processed
all the macro procedures and generated corresponding PL/I templates.
We then translated these templates into Java code using two LLMs:
Llama-4-Scout-17B-16E-Instruct~\cite{llama4scout} and Mistral-Small-3.2-24B-Instruct-2506~\cite{mistral2025}.
All templates were successfully translated into the expected Java methods.
\begin{figure}[t]
\begin{lstlisting}
 {_V_} = 10{#if(#1)} + {_X_}{#endif};
\end{lstlisting}
\begin{lstlisting}
@PLIVariable(name="v", value="{_V_}", type="int")
@PLIVariable(name="x", value="{_X_}", type="int")
public void targetMethod() {
  if (x1) { setV(10 + getX()); }
  else { setV(10); }
}
\end{lstlisting}
\vspace{-5pt}
\caption{An Example of an Unbalanced Template}
\label{fig:unbalanced}
\vspace{-10pt}
\end{figure}
Even in cases where the conditional constructs in the template and the sentences generated
by the macro procedure are unbalanced, as shown in Figure~\ref{fig:unbalanced},
the translation was successfully achieved.
Such macro procedures are commonly found in real-world code,
and this highlights one of the advantages of using LLMs.

\begin{figure}[t]
\begin{lstlisting}
static String updateMax(String prefix, String max, String x1, String x2) {
  StringBuilder s = new StringBuilder();
  if (x2.isEmpty()) {
    s.append("if (").append(prefix).append(x1).append(" >= ")
     .append(prefix).append(max).append(") {\n");
    s.append("    ").append(prefix).append(max).append(" = ")
     .append(prefix).append(x1).append(";\n");
    s.append("}\n");
  } else {
    ...
  }
  return s.toString();
}
\end{lstlisting}
\vspace{-5pt}
\caption{Java Code Generated by Llama-4-Scout}
\label{fig:generated:llama}
\vspace{-10pt}
\end{figure}

For comparison with the naive direct application of LLMs, we conducted an additional experiment
in which we asked the LLMs to translate the original PL/I macro procedures directly into Java.
However, none of the macro procedures were translated appropriately, despite experimenting
with several prompt patterns.
For example, Figure~\ref{fig:generated:llama} shows the Java method generated by Llama-4-Scout.
It is a piece of code that constructs another Java code, but the LLM was unable to generate
the actual target code further.
We observed similar results for the other macro procedures.

\section{Future Plans}
\label{sec:plans}

There are two primary directions planned for future work.
\\
\noindent
\textbf{Expansion of Benchmarks:}
In the current experiments, the benchmark dataset was constructed based on a limited set of
patterns of PL/I macro procedures derived from real-world code.
In future work, we plan to investigate a broader range of patterns in actual code and
expand the set of macro procedures targeted for transformation.
Such datasets will provide more practical insights.
\\
\noindent
\textbf{Generation of More Natural Java Code:}
This paper restricts the patterns of the target PL/I macro procedures to those that can
be expressed using Java annotations.
Java annotations help simplify code by abstracting complex logic and improves modularity allowing
developers to focus on core functionality.
However, annotations can introduce a learning cost and make the underlying behavior less transparent.
The plain key-value-based Java code generated by the annotation processor often suffers
from poor readability and runtime performance overhead.
Therefore, we plan to explore techniques for further generating annotation-free and human-readable Java code.

\section{Conclusion}
\label{sec:conclusion}

Translating PL/I macro procedures into Java poses significant challenges
due to their nature as code that generates code.
Direct translation using LLMs often fails to capture the semantics of these macros.
The templatization approach presented in this paper offers a robust solution
that preserves code modularity and maintainability.
We validated this approach through experiments on ten representative macro procedure patterns,
all of which were successfully translated into Java using our method.

\bibliographystyle{ACM-Reference-Format}
\bibliography{references}

\end{document}